\documentclass[a4paper,11pt]{article}
\usepackage{pos}

 \usepackage{bbold}
 
 \usepackage{caption}
\usepackage{subcaption}
\usepackage{slashed}
\usepackage{amsmath}    
\usepackage{dsfont}

\begin{document}
\title{The quark propagator and quark-gluon vertex from lattice QCD at finite temperature}
\ShortTitle{Quark propagator and
  quark-gluon vertex from LQCD at finite T}

\author*[a,b]{Jesuel Marques}
\author[b]{Gerhard Kalusche}
\author[a]{Tereza Mendes}
\author[c]{Paulo J. Silva}
\author[b]{Jon-Ivar Skullerud}
\author[c]{Orlando Oliveira}

\affiliation[a]{Instituto de Física de São Carlos, Universidade de São Paulo,\\ C.P. 369, 13560-970, São Carlos, SP, Brazil}
\affiliation[b]{Department of Theoretical Physics, Science Building, Maynooth University, Maynooth, Co. Kildare, W23 F2H6, Ireland}
\affiliation[c]{CFisUC, Department of Physics, University of Coimbra, 3004-516 Coimbra, Portugal}

\emailAdd{jesuel.leal@usp.br}
\emailAdd{gerhard.kalusche.2021@mumail.ie}
\emailAdd{mendes@ifsc.usp.br}
\emailAdd{psilva@uc.pt}
\emailAdd{jonivar@thphys.nuim.ie}
\emailAdd{orlando@fis.uc.pt}

\abstract{The quark-gluon vertex is an important object of QCD. Studies have shown that this quantity is relevant for the dynamical chiral symmetry breaking pattern in the vacuum. The goal of our project is to obtain the quark-gluon vertex at finite temperature around the deconfinement/chiral transition using the tools provided by lattice QCD. It will be the first time that the quark-gluon vertex at finite temperature is determined using lattice QCD. The propagators, which are a by-product of this project, are also of interest in themselves. The configurations used were generated by the FASTSUM collaboration. In this contribution, we describe our motivations and goals, some technical details of the determination and report on the status of the calculation.}

\FullConference{%
  The 39th International Symposium on Lattice Field Theory (Lattice2022),\\
  8-13 August, 2022 \\
  Bonn, Germany 
}


\maketitle

\section{Introduction and aim of the project}
 The aim of this project is to provide a non-perturbative determination of the quark-gluon vertex at finite temperature. It is expected that Quantum Chromodynamics (QCD) should provide a first-principles theoretical explanation for the phenomena observed in the hadronic world. Two of these are: the spontaneous chiral symmetry breaking, responsible for the dynamically generated mass of hadrons; and confinement, the observation that quarks and gluons are not detected outside of their bound states. Both phenomena have eluded a satisfactory detailed theoretical explanation since the 70s and are still important themes under discussion in the strong interaction community. 
 
 The effects of non-perturbative phenomena must appear in the N-point functions of the theory, since they completely determine the theory. The quark-gluon three-point function describes the interaction between the quark and the gluon fields. An approach suggesting a relationship between Green's functions and non-perturbative phenomena is the Gribov-Zwanziger scenario \cite{Gribov:1977wm,Vandersickel:2012tz}, which predicts that confinement is detectable in the 2-point functions (propagators), through a violation of positivity of the associated spectral function, and a specific behaviour of these functions for momenta in the infrared region. In the context of the Schwinger-Dyson equations approach, the non-perturbative non-Abelian contribution to the full quark-gluon vertex is needed to account for the observed chiral symmetry breaking pattern, since the bare or the Abelian vertices are not sufficient \cite{Kizilersu:2021jen}. In fact, as has also been noted in \cite{Kizilersu:2021jen}, several studies following the approach of Schwinger-Dyson equations explored the effect of the non-leading form factors of the quark-gluon vertex on chiral symmetry breaking and hadron phenomenology.

In search of insights to help understand the mechanisms of chiral symmetry breaking and confinement, high temperature may be a worthwhile path and an interesting laboratory. This is so because, in this regime, there occurs a transition to a state called ``quark-gluon plasma'' (QGP), in which the particles of the theory are not confined to their bound states \cite{Collins:1974ky,Cabibbo:1975ig}, and also because at high temperatures restoration of chiral symmetry takes place. One might think that non-perturbative effects should be absent in hot QCD, since the typical energy scales are also high and this would allow perturbation theory to be applicable. Actually, as evidenced by the fact that the QGP appears to behave as a nearly perfect liquid, QCD seems to be strongly coupled at least up to temperatures as high as 450 MeV. The magnetic sector is also known to be non-perturbative for all $T$. In order to have a correct picture of deconfinement and chiral restoration transitions and a description of the QGP, one must take into account non-perturbative effects. The extension of the lattice QCD framework to study the high temperature regime is straightforward, which makes it well suited to research into this topic.

Since studies show that the dynamics of the quark-gluon 3-point function are so intimately connected with chiral symmetry and since chiral symmetry is expected to be restored at high temperatures, it certainly would be quite useful to have an independent lattice determination of this object at finite temperature. These results could also be used to cross-check or validate the assumptions used by other methods.

The quark propagators, which are also of interest in themselves, are a subproduct of this project, because we need first to determine them in order to calculate the quark-gluon vertex. Quark propagators and the quark-gluon vertex functions on the lattice have been studied by some of the authors in \cite{Kizilersu:2021jen,Skullerud:2001aw,Oliveira:2019erx,Skullerud:2002ge,Skullerud:2003qu,Skullerud:2004gp,Oliveira:2018lln}. In general it is observed that quark propagators exhibit dynamical chiral symmetry breaking by an enhancement of the running mass function in the infrared region to a value of $\approx 350\textrm{MeV}$, while the wave-function renormalization $Z$ is suppressed with respect to its tree-level value \cite{Skullerud:2001aw,Oliveira:2018lln}. At finite temperature and for quenched gauge configurations \cite{Oliveira:2019erx}, the authors found significant differences between the propagator's form factors above and below the transition temperature. In the review \cite{Fischer:2018sdj}, the authors show how the Dyson-Schwinger approach is able to reproduce lattice results in the vacuum for Landau gauge and they also predict a critical scaling law for the scalar form factor with temperature, at least close to the chiral limit with two flavours. Using the fRG approach, the authors of \cite{Gao:2021wun} were also able to reproduce lattice results for the quark propagator in Landau gauge at zero temperature.

The study of the quark-gluon vertex on the lattice in the vacuum pursued in \cite{Kizilersu:2021jen} found that the form factors which characterize the vertex all have significant strengths and that they are enhanced at low momenta when one considers two flavors of dynamical quarks in comparison to non-dynamical quarks (quenched approximation). This enhancement goes beyond the one-loop prediction by many orders of magnitude, highlighting the importance of non-perturbative contributions. In \cite{Skullerud:2003qu}, besides a negligible mass dependence, substantial deviations from the abelian form were found for soft gluon kinematics. At the quark-reflection point and for low energies, the form factor related to the chromomagnetic moment was found to contribute significantly to the interaction strength. The enhancement in the infrared was observed to be a general feature of the vertex irrespective of the kinematics in \cite{Skullerud:2004gp}. Dyson-Schwinger studies in the vacuum \cite{Alkofer:2013qoc} found that, for symmetric momenta, both vertex form factors that are 0 at tree level have similar strengths once one goes into the non-perturbative regime. For the symmetric kinematical point, the authors of  \cite{Gao:2021wun} see an enhancement of all form factors in a region of momenta between 0.5 and 1 GeV, which they claim to be necessary to obtain the correct amount of chiral symmetry breaking from the gap equation.

\section{Technical details of the lattice action}
This project uses anisotropic lattices, which are characterized by a different lattice spacing in the spatial and temporal directions, and which allow a good scan of the pseudocritical temperature region. We use the ensembles Gen2 and Gen2L generated by the FASTSUM collaboration \cite{Aarts:2007pk, Aarts:2014nba, Aarts:2020vyb}. This approach has already been used to calculate the spectral functions of charmonium at high temperature \cite{Aarts:2007pk}, transport properties of the quark-gluon plasma \cite{Aarts:2014nba}, and also quantities characterizing the QCD thermal transition, such as the chiral condensate and various susceptibilities \cite{Aarts:2020vyb}, just to mention a few applications.

A Symanzik improved action is used for the gluon sector. The action for the dynamical Wilson fermions  ($N_f=2+1$) with clover improvement is $ S_F=\sum_{x,y}\bar{\psi}_x \mathcal{M}_{xy} \psi_y$,
\begin{equation}
    \mathcal{M}_{xy}=\delta_{x,y}\left[\left(a_t m_0+1+\frac{3}{\gamma_f}\right)\, -\frac{c_t}{2}\sum_i \sigma_{i4}F_{x;\;i4}-\frac{c_s}{2}\sum_{i<i'} \sigma_{ii'}F_{x;\;ii'}\right]-\frac{1}{\gamma_f}\sum_{i=1}^3 H_{xy;\;i}-H_{xy;\;4}
    \label{eq:diracwilsonanisotropicoperator}
\end{equation}
where $H_\mu$ are hopping terms
\begin{equation}
    H_{xy;\;\mu}=\frac{1}{2}\left(\mathds{1}_s-\gamma_\mu\right)U_{x, \mu}\delta_{x+\hat{\mu},\,y}+\frac{1}{2}\left(\mathds{1}_s+\gamma_\mu\right)U_{y, \mu}^\dagger\delta_{x-\hat{\mu},\,y}.
\end{equation}

The relevant parameters for this article are given in Table \ref{tab:parameters}. The relationship between the spatial and temporal spacing is usually expressed as $a_s=\xi a_t$. The physical parameters for the Gen2 ensemble are: $a_s=0.1205(8)\textrm{fm}$,\, $\xi=3.444(6)$ and $m_\pi=384(4)\textrm{MeV}$ and for Gen2L: $a_s=0.1136(6)\textrm{fm}$,\, $\xi=3.453(6)$ and $m_\pi=236(2)\textrm{MeV}$. The strange quark mass has its physical value. We have access to configurations with spatial volume of $24^3$ and $32^3$, and we use the fixed scale approach to vary the temperature, meaning that $a_t$ is kept constant and we change the extent of the lattice in the imaginary time direction.

\begin{table}[h!]
\centering
\begin{tabular}{|c|c|c|c|}
\hline
     $\gamma_f$ & $c_s$  & $c_t$ & $a_t m_0$ (Gen2, Gen2L)\\
     \hline
       3.399   &  1.5893 & 0.90278 & -0.0840, -0.0860 \\
     \hline
\end{tabular}
\caption{Bare fermion anisotropy $\gamma_f$, spatial and temporal clover coefficients $c_s$ and $c_t$, and bare light quark masses in units of the temporal lattice spacing $a_tm_0$, used by FASTSUM to generate the configurations used in this project, as tuned by HadSpec \cite{HadronSpectrum:2008xlg}.}
\label{tab:parameters}
\end{table}
\section{Gauge-fixing}
\label{sec:gaugefixing}

Green's functions are gauge-dependent quantities, and therefore one needs to choose a gauge to obtain well-defined results. In the vacuum, a popular choice is Landau gauge; however, for the case at hand, with anisotropic lattices and due to the breaking of Lorentz symmetry caused by temperature, Coulomb gauge turns out to be more clearly defined. This opens up the possibility for exploring qualitative or quantitative differences between Coulomb and other gauges. Coulomb gauge-fixing is implemented on the lattice by extremizing the functional
\begin{equation}
    W[U; g]=\sum_x\sum_{\mu=1}^3 \textrm{Re}\textrm{Tr} \left[U_{x,\mu}^g\right],
\end{equation}
as a function of all possible gauge-transformations $g=\{g_x\}$, with $U=\{U_{x, \mu}\}$ fixed, for the configurations of the ensemble. The gauge-transformed links are $U_{x,\mu}^g=g_x U_{x,\mu} g_{x+\hat{\mu}}^\dagger$. This extremization procedure can be shown to be equivalent to the Coulomb gauge condition, $\vec{\nabla}\cdot \vec{A}=0$, on the lattice, where $A_{x,\mu}$ is the four-vector potential \cite{Cucchieri:2003fb}. The discretization of $A_{x,\mu}$ is the usual $A_{x,\mu}=(U_{x,\mu}-U_{x,\mu}^\dagger)/2 i-\textrm{trace}$.

The algorithm chosen to perform the gauge-fixing is overrelaxation \cite{Mandula:1990vs}, which has good scaling properties, in comparison to other algorithms\footnote{For more information about the gauge-fixing algorithm and its scaling, see \cite{Suman:1993mg, Leal:2022ojc}.}. In order to monitor the convergence of the gauge-fixing, we use
\begin{equation}
e=\frac{1}{|\Lambda|}\sum_{x} \sum_{b=1}^{N_c^2-1}\left(\vec{\nabla}\cdot \vec{A_x^b}\right)^2,
\end{equation}
where $|\Lambda|$ is the number of sites on the lattice and $A_{i, x}^b$ is defined as the color components $A_{x,i}=\sum_{b=1}^8 A_{x,i}^b \frac{\lambda_b}{2}$, $\lambda_b$ being the Gell-Mann matrices. The gauge-fixing program sweeps through the lattice and periodically measures the value of $e$, stopping when it is below a certain tolerance, which we fix at $10^{-16}$ in this study.

An openMP parallelized C implementation for the gauge-fixing procedure was used \cite{gaugefixingcoulomb}. Parallelization is made possible by a checkerboard subdivision of the lattice, which allows one to slice the lattice and perform gauge fixing on each slice by a different processor core. One then takes turns in updating either the even or odd sites. In this way a speedup factor almost proportional to the number of cores can be obtained, since the serial parts of the code are minimal. The gauge-fixing procedure takes about 6 minutes on average on a 4-core Intel i3-10100 processor for a $16^3\times128$ configuration.

\section{Preliminary results}
We first determine the quark propagators, which are best represented in momentum space. Thus, after performing the inversion\footnote{USQCD-Chroma \cite{Edwards:2004sx} has been used to perform the inversions.} of $\mathcal{M}_{xy}$ (Eq.\ \ref{eq:diracwilsonanisotropicoperator}), we apply a Fourier transformation.  From symmetry considerations, the inverse quark propagator in momentum space on an anisotropic lattice should have the form
\begin{equation}
    S^{-1}(a_t P_4, a_s \vec{P})=Z(a_t P_4, a_s \vec{P})\left[\frac{ia_s \vec{\gamma}\cdot\vec{K}(aP)}{\gamma_f}+a_t M(a_t P_4, a_s \vec{P})+ i a_t K_4(aP) \gamma_4 \sigma(a_t P_4, a_s \vec{P})\right],
    \label{eq:invpropgenericformanisotropic}
\end{equation}
where $a_\mu K_\mu(aP)=\sin\left(a_\mu P_\mu\right)$. By taking traces over this expression and combining them, we can obtain the $M$, $Z$ and $\sigma$ form factors.

The spatial and temporal components of the fermionic momenta allowed by the lattice spacing and lattice extent are 
\begin{equation}
a_s P_i = \frac{2\pi}{N_i} n_i,\quad a_t P_4 = \frac{2\pi}{N_4}\left(n_4+\frac{1}{2}\right),
\end{equation}
the difference being due to the fermionic antiperiodic boundary condition in the time direction. The lattice extent in direction $\mu$ is $N_\mu$, and $1 \leq n_\mu \leq N_\mu$. Since we are interested more in the infrared than in the ultraviolet, we calculate only momenta corresponding to $1\leq n_\mu \leq N_\mu/4$. We also average over momenta related by permutations in the spatial directions.

\subsection{Uncorrected results}
\begin{figure}[ht!]
    \centering
    \includegraphics[scale=0.395]{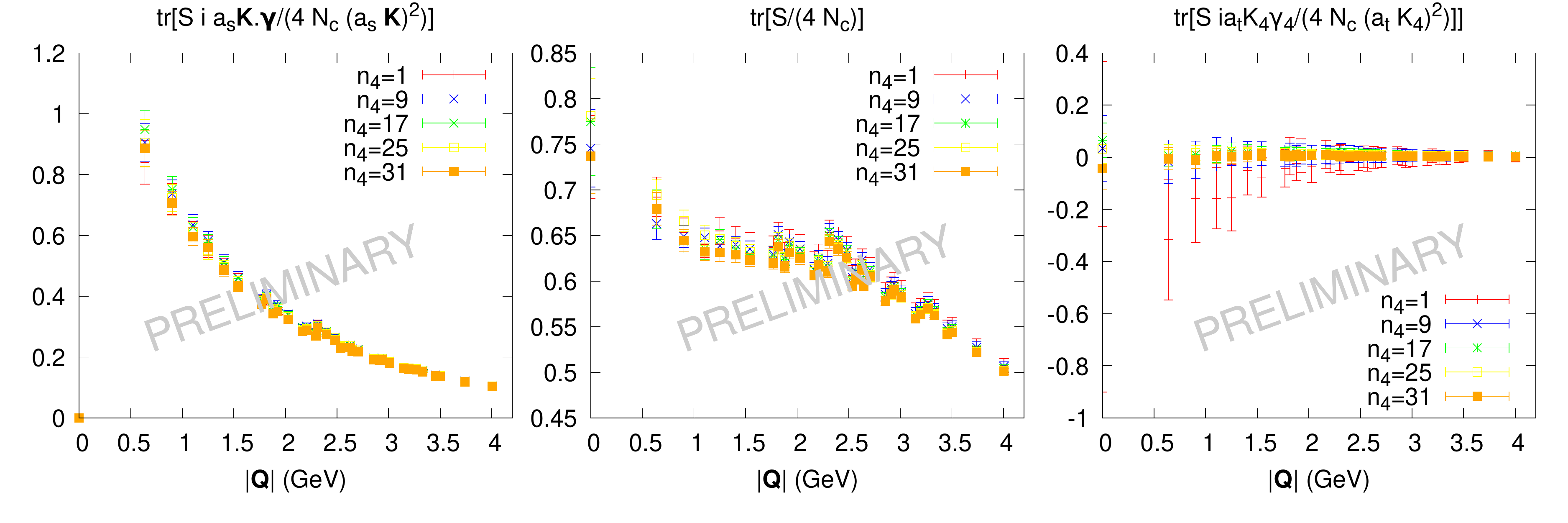}
     \caption{Traces for 51 configurations on a $16^3\times128$ lattice for different energies $p_4$ as a function of the norm of the spatial lattice momentum $\vec{Q}$. A selection of energies was made }
    \label{fig:nonaveraged_traces}
\end{figure}

Here we show preliminary results for zero temperature, using a lattice with size $16^3\times128$. We found no significant dependence on the energy $p_t$, for any of the traces of the propagator, as Fig.\ref{fig:nonaveraged_traces} shows as a function of the norm of the spatial part of the lattice momentum $ Q_\mu(aP)=\frac{2}{a_\mu}\sin\left(a_\mu P_\mu/2\right)$.

\begin{figure}[ht!]
    \centering
    \includegraphics[scale=0.55]{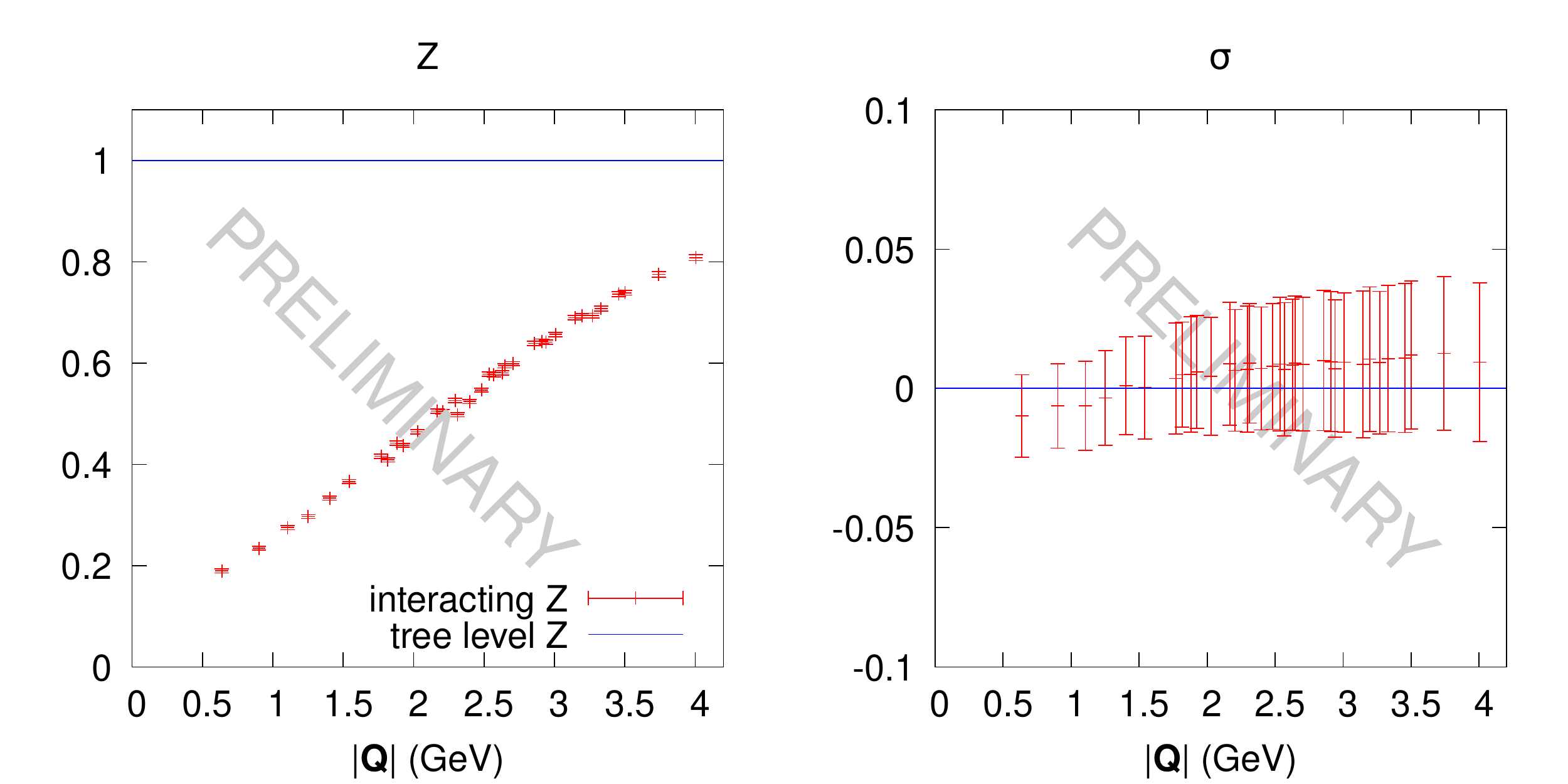}
     \caption{Unrenormalised Z and $\sigma$ form factors from 51 configurations on a $16^3\times128$ lattice.}
    \label{fig:Z&sigmaplots}
\end{figure}

 We then chose to average these quantities over $p_t$ and plot the form factors calculated from them. For the unrenormalised $Z$ and $\sigma$ form factors, results are shown in Fig. \ref{fig:Z&sigmaplots}. The Z form factor is suppressed in the infrared and increases with increasing momentum. The blue line shows the value Z=1 for the free quark propagator. We see that the $\sigma$ form factor is very small for all momenta. The energy independence of the form factors and $\sigma$ being small are in agreement with what was found in \cite{Burgio:2012ph} using staggered fermions. 

\begin{figure}[ht!]
    \centering
    
    \includegraphics[scale=0.55]{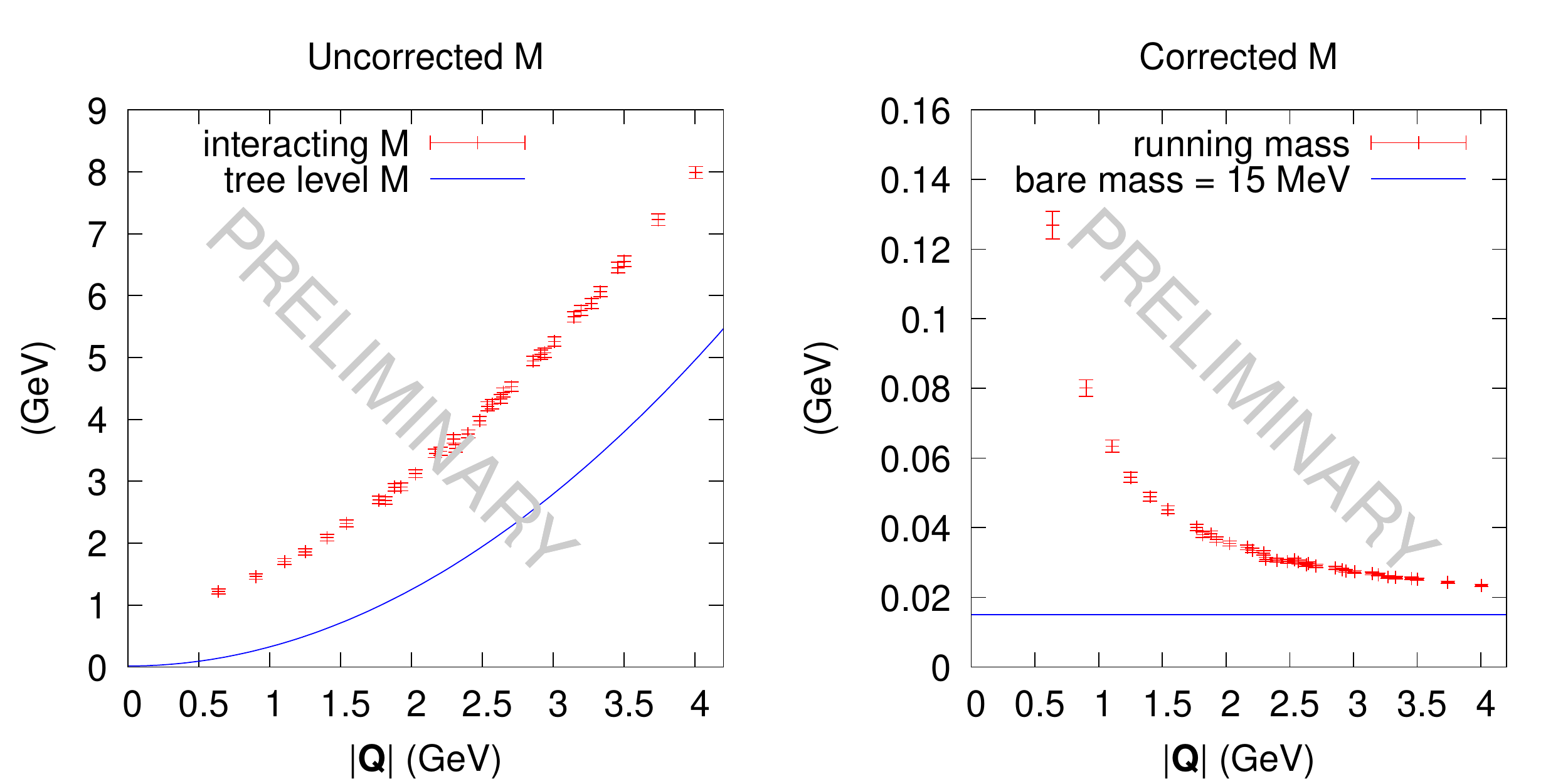}
     \caption{Uncorrected and corrected M form factor from 51 configurations on a $16^3\times128$ lattice.}
    \label{fig:Mplot}
\end{figure}

Fig. \ref{fig:Mplot} shows the results for the M form factor. The red points on the plot on the left show the values as extracted from the traces of the propagator. The tree-level expression for Wilson-fermions is shown as a blue curve. In order to obtain reasonable results, one needs to correct for Wilson-fermion lattice artifacts as explained in Section \ref{sec:corrections}.

\subsection{Corrections}
\label{sec:corrections}

One well-known drawback of Wilson type fermions is that they explicitly break chiral symmetry on the lattice. In order to determine the bare quark mass, one needs to correct for this, by subtracting the critical mass from the bare mass appearing on the action
\begin{equation}
    a_t\,m=a_t\,m_0 - a_t\, m_c.
\end{equation}
The critical mass is the value of $a_t\,m_0$ for which the pion mass vanishes. Using data from \cite{HadronSpectrum:2008xlg}, assuming a relation $(a_t\,m_\pi)^2 \propto a_t\,m$, we obtain  $a_t\,m_c=-0.0866(6)$ by extrapolating the result of the fit to $a_t\,m_\pi=0$. This gives  $a_t\,m = 0.0026(6)$, or $m = 15(4)\textrm{MeV}$, using the value of $a_t\,m_0$ for our ensemble.

Another drawback that Wilson fermions suffer from is the fact that its mass form factor $M$ has a momentum dependence even for free quarks (i.e.  when $U_{x,\mu}=\mathds{1}_c$),
\begin{equation}
    a_t M^\textrm{free}(aP)=a_tm+\frac{1}{2\gamma_f}\left(a_s\vec{Q}(aP)\right)^2+\frac{1}{2}\left(a_t Q_4(aP)\right)^2,
    \label{eq:freemass4d}
\end{equation}
whereas the continuum mass would simply be $a_tm$ for a free theory. In order to get sensible results, this effect needs to be corrected. We choose the so called multiplicative correction \cite{Skullerud:2001aw}
\begin{equation}
    a_t M^\textrm{corrected}(aP)=a_t\, m  \frac{a_t M(aP)}{ a_t M^\textrm{free}(aP)}.
\end{equation}
However, since our data does not show any energy dependence, we cannot use Eq.\ \ref{eq:freemass4d}, as this would introduce a spurious energy dependence to our results. Instead, as the Coulomb gauge-fixing seems to restrict the dependence to the three-momentum, we use a spatial three-dimensional free propagator for the correction. In momentum space, restricted to the three dimensions of each time slice of the anisotropic lattice, the Wilson propagator mass function, for the free case is

\begin{equation}
    a_t M^\textrm{free}_{ 3d}(aP)=a_tm+\frac{1}{2\gamma_f}\left(a_s\vec{Q}(aP)\right)^2.
    \label{eq:freemass3d}
\end{equation}

Qualitatively, the corrected data, shown in the right hand side of Fig.\ \ref{fig:Mplot} shows the expected form for the running mass, which approaches the bare mass for high momentum and increases for low momentum, displaying what we interpreted as a manifestation of the dynamical chiral symmetry breaking in the infrared. The qualitative suppression of the Z form factor in the infrared was also obtained in \cite{Burgio:2012ph}. However, quantitatively, the results for the mass form factor differ from those obtained by \cite{Burgio:2012ph} by a factor of approximately 2 for the lowest momentum data point, where staggered fermions with a bare mass of of $15.7 \textrm{ MeV}$, similar to ours, were used. We are currently performing further checks to try to understand the source of this discrepancy.

\section{Conclusion and next steps}
As part of our project to obtain the quark-gluon vertex on the lattice at finite temperature, we obtained the quark propagators at zero temperature on Coulomb gauge. Our analysis shows the expected manifestation of dynamic chiral symmetry breaking for the mass function form factor.

\subsection{Improvement and rotation}
For off-shell quantities, such as the propagators, in principle one needs to improve the fields in tandem with the clover action improvement in order to obtain $\mathcal{O}(a^2)$ correct results \cite{Skullerud:2001aw,Dawson:1997gp,Capitani:2000xi}. This implies that the naïve propagators, obtained by a plain inversion of the Dirac-Wilson operator, $S_\textrm{naïve\;xy}=\mathcal{M}_{xy}^{-1}$, need to be substituted by
\begin{equation}
    S_{R\;xy}=\left\langle\left[\delta_{x,w}-\frac{\slashed{D}_{xw}}{4}\right]\left(1+\frac{a_tm}{2}\right)S_{\textrm{naïve}\;wz}\left[\delta_{z,y}+\frac{\overleftarrow{\slashed{D}}_{zy}}{4}\right]\right \rangle_U,
    \label{eq:rotatedpropdefinition}
\end{equation}
which we refer by the name of rotated propagator, rotation being the action of the operator in square brackets. Our rotated results are currently being analyzed.
\subsection{High-temperature and vertex}
We can extend our study straightforwardly to high temperatures just by using the Gen2 configurations with a shorter imaginary time extent. Work is in progress in analyzing the results for high temperature.

Having the quark and gluon propagators ($S(p)$ and $D(q)_{\mu\nu}$, respectively) and the Fourier transformed four-potential fields $A_\mu(q)$ we can calculate the amputated vertex by
\begin{equation}
    \Lambda_\mu^{a}(p, q)=S^{-1}(p)V_\nu^a(p, q)S^{-1}(p+q)D^{-1}(q)_{\nu\mu}
\end{equation}
with
\begin{equation}
    V_\mu^a(p, q)=\left\langle S(p; U)A_\mu^a(q)\right\rangle_U.
\end{equation}

We shall calculate it for a selection of kinematic configurations, such as the soft gluon limit ($q\rightarrow 0$), quark hard reflection ($k_\mu=-p_\mu=q_\mu/2$, with $k=p+q$) and the orthogonal kinematics ($q\cdot(p+k)=0$ and $k^2=p^2$). The quark hard reflection allows us to study the chromomagnetic form factor, associated to the $\sigma_{\mu\nu}q_\nu$ basis element, and the orthogonal kinematics, the rank 3 tensor structure which accompanies $[\slashed{p},\slashed{k}]\gamma_\mu$. One complication that we anticipate is that the tensor structure at finite temperature is more involved than the one for the vacuum \cite{Ayala:2001mb}, and may need to change further to accommodate effects of the Coulomb-fixing and anisotropic lattices, as happened to the quark propagator.

\acknowledgments

Jesuel Marques has been supported by grants \#2019/10913-0 and \#2021/11101-9 of the São Paulo Research Foundation (FAPESP). Paulo Silva and Orlando Oliveira are supported by FCT contracts UIDB/04564/2020 and UIDP/04564/2020. Paulo Silva is also supported by FCT contract CEECIND/00488/2017. Gerhard Kalusche was supported by a Maynooth University SPUR scholarship. We acknowledge computing time provided for this project by the Irish Centre for High-End Computing (ICHEC).

\end{document}